\documentclass[
    ,final            % use final for the camera ready runs
    ,tnotealph
  ]
  {aipproc}

\layoutstyle{6x9}

\begin{document}

\title{Abundance analysis of DAZ white dwarfs}


\author{Ad\'ela Kawka}{
  address={Astronomick\'y \'ustav, Akademie v\v{e}d \v{C}esk\'e republiky, 
  Fri\v{c}ova 298, CZ-251 65 Ond\v{r}ejov, Czech Republic}
}

\author{St\'ephane Vennes}{
  address={Astronomick\'y \'ustav, Akademie v\v{e}d \v{C}esk\'e republiky, 
  Fri\v{c}ova 298, CZ-251 65 Ond\v{r}ejov, Czech Republic}
}

\author{Franti\v{s}ek Dinnbier}{
  address={Faculty of Mathematics and Physics, Charles University, 
  V Hole\v{s}ovi\v{c}k\'ach 2, CZ-180 00 Praha, Czech Republic}
}

\author{Helena Cibulkov\'a}{
  address={Faculty of Mathematics and Physics, Charles University, 
  V Hole\v{s}ovi\v{c}k\'ach 2, CZ-180 00 Praha, Czech Republic}
}

\author{P\'eter N\'emeth}{
  address={Astronomick\'y \'ustav, Akademie v\v{e}d \v{C}esk\'e republiky, 
  Fri\v{c}ova 298, CZ-251 65 Ond\v{r}ejov, Czech Republic}
}

\begin{abstract}
We present an abundance analysis of a sample of 33 hydrogen-rich (DA) white dwarfs. We have
used archival high-resolution spectra to measure abundances of calcium, 
magnesium and iron in a set of 30 objects. In addition, we
present preliminary calcium abundances in three new white dwarfs based on low-dispersion
spectra. We investigate some abundance ratios (Mg$/$Ca, Fe$/$Ca) 
that may help uncover the composition of the accretion source.
\end{abstract}

\maketitle

%%%%%%%%%%%%%%%%%%%%%%%%%%%%%%%%%%%%%%%%%%%%
%% MAINMATTER
%%%%%%%%%%%%%%%%%%%%%%%%%%%%%%%%%%%%%%%%%%%%

\section{Introduction}

Most white dwarfs possess a hydrogen-rich atmosphere (DA), with 
the remainder having a helium-rich atmosphere. 
Heavy elements are expected to sink below the photosphere because of the
high surface gravity leaving either a pure-hydrogen or pure-helium atmosphere. However,
a significant fraction of white dwarfs show traces of elements heavier
than helium. In hot white dwarfs ($> 20\,000$ K) radiative levitation can
support metals against gravitational settling \citep{cha1995}. But in
cooler white dwarfs an external source is required to feed the atmosphere with
heavier elements. Scenarios explaining the presence of
these elements in white dwarf atmospheres include accretion from the
interstellar medium \citep{dup1993,koe2006}, from unseen companions \citep{deb2006},
or from asteroidal/planetary material in circumstellar debris discs 
\citep{gra1990,deb2002,jur2003}.

Approximately one quarter of cool DA white dwarfs show the presence of
metals in their spectra \citep[DAZ;][]{zuc2003}, and out of these approximately 20\%
have a mid-infrared excess that is consistent with a circumstellar disc 
\citep{far2009}. Preliminary population statistics \citep{far2009} showed that 
white dwarfs with a higher calcium abundance were more likely to show IR excess
than those with a lower abundance. Using Spitzer observations of several
DAZ white dwarfs and their helium-rich counterparts (DZ), \citet{far2010} showed 
that the size of a debris disc can vary, and that 
some debris discs may be narrow enough to remain undetected.

We present abundance measurements for a sample of known and new DAZ white 
dwarfs. The analysis of known DAZ white dwarfs was prompted by the discovery
of magnesium and other heavy elements in the heavily polluted white dwarf GALEX~J1931+0117 \citep{ven2010} and
the lack of detection of metals other than calcium in other DAZ white dwarfs.
We therefore initiated a search for heavy elements in DAZ white dwarfs
using archival and new spectroscopy. The new DAZ white dwarfs
were identified as part of our spectroscopic observations of white dwarf
candidates in high-proper motion catalogues (NLTT, LSPM).

\section{Atmospheric properties}

\begin{figure}
\includegraphics[viewport=1 240 550 530, clip, width=0.70\textwidth]{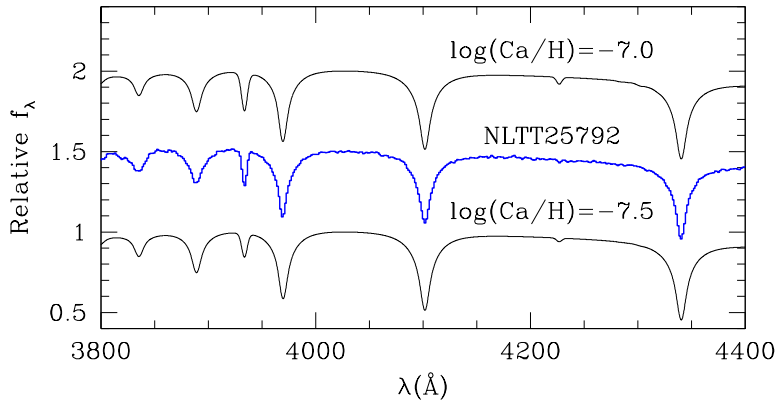}
\caption{FORS1 spectrum of NLTT25792, showing the Ca\,{\sc ii} $\lambda$3933.7 \AA\ and
Ca\,{\sc i} $\lambda$4226.7 \AA\ lines, compared to synthetic spectra at $T_{\rm eff} = 8160$ K, $\log{g} = 8$ and $\log{\rm (Ca/H)} = -7$ and $-7.5$.}
\label{fig_nltt25792}
\end{figure}

We have extracted UV-Visual Echelle Spectrograph (UVES) and High Resolution Echelle Spectrograph (HIRES) exposures 
for a sample of DAZ white dwarfs from the
European Southern Observatory (ESO) and the Keck archives, respectively.
We have also identified three new DAZ white dwarfs in the New Luyten Two-Tenths (NLTT) catalogue. The low
resolution spectra obtained using the 4m telescope at Cerro Tololo 
Inter-American Observatory (CTIO) or using the VLT-Kueyen telescope and FORS1
showed that NLTT~6390, NLTT~23966 and NLTT~25792 are cool DAZ white dwarfs.
Calcium in the new DAZ white dwarfs is clearly photospheric and not interstellar.

We determined the effective temperature and surface gravity of the white 
dwarfs by fitting the Balmer line spectra with synthetic spectra for pure hydrogen models.
\citet{kaw2006} summarizes the model atmosphere calculations.
The grid covers effective temperatures from 4500 K to 50\,000 K and surface
gravities from $\log{g} = 5.75$ to $9.50$. The model spectra were convolved with the
instrumental resolution. The UVES spectra were binned to 1\AA\ before fitting
the Balmer lines with model spectra.
Because the HIRES echelle spectra are not flux calibrated, we 
obtained a Sloan Digital Sky Survey (SDSS) spectrum for WD1455+298 and fitted the Balmer lines to determine
the effective temperature and surface gravity. The atmospheric parameters
for WD0208+396 were taken from \citet{gia2005}.

We searched for the presence of metals in the high-resolution spectra. Several
lines of calcium, magnesium and iron were detected, but in the present analysis
we restricted the abundance determinations to 
Ca\,{\sc ii} $\lambda$3934 \AA, Mg\,{\sc i} $\lambda$5184 \AA, Mg\,{\sc ii} $\lambda$4481 \AA, 
and Fe\,{\sc i} $\lambda$3581\AA. We calculated a set of LTE spectra with varying 
calcium, magnesium and iron
abundances at $\log{g} = 8$ and with temperatures ranging from 6000 
to 25\,000 K. We then computed synthetic equivalent widths for the selected lines 
within windows of $\Delta = 10$\AA\ for Ca\,{\sc ii} and Mg\,{\sc ii}, $\Delta = 8$\AA\ for Mg\,{\sc i}, and 
$\Delta = 6$\AA\ for Fe\,{\sc i}. 

For NLTT~25792, we estimated the calcium abundance by fitting synthetic spectra to the 
Ca\,{\sc ii}$\lambda$3934\AA\ line profile.
Figure~\ref{fig_nltt25792} shows the observed FORS1 spectrum compared to
two model spectra with $\log{\rm (Ca/H)} = -7.0$ and $-7.5$.

For the remaining stars we determined the abundances by fitting the
measured equivalent widths with the synthetic equivalent widths.
Since we did not have a spectrum covering Ca\,{\sc ii} H and K lines for
WD0208+396 we used the abundance determined by \citet{zuc2003}. The
atmospheric parameters and abundances for GALEX~J1931+0117 were taken from 
\citet{ven2010}.

\begin{figure}
\includegraphics[width=0.65\textwidth]{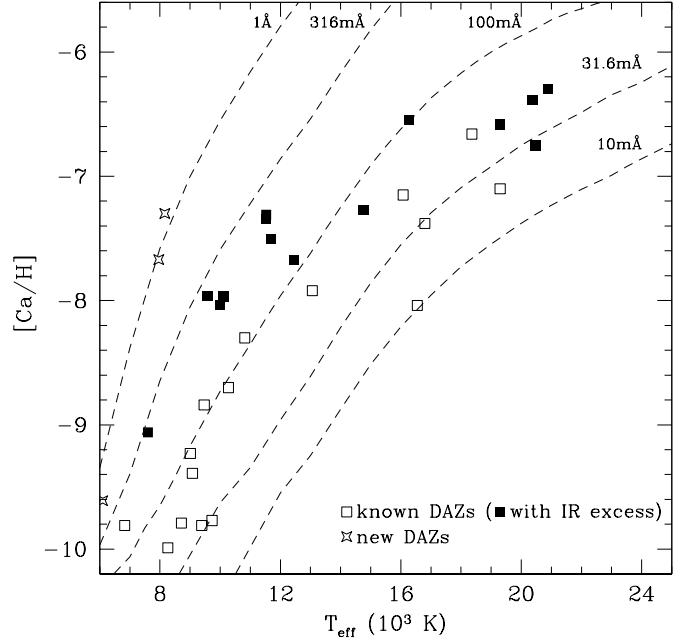}%
\caption{Abundance of calcium versus effective 
temperature. Filled squares are white dwarfs with confirmed infrared excess, 
and open squares are white dwarfs either without infrared excess or that have 
not been checked for infrared excess. 
The dashed lines are curves of constant equivalent 
widths for Ca\,{\sc ii} $\lambda$3933\AA.}
\label{fig_abunca}
\end{figure}

\begin{figure}
\includegraphics[width=0.65\textwidth]{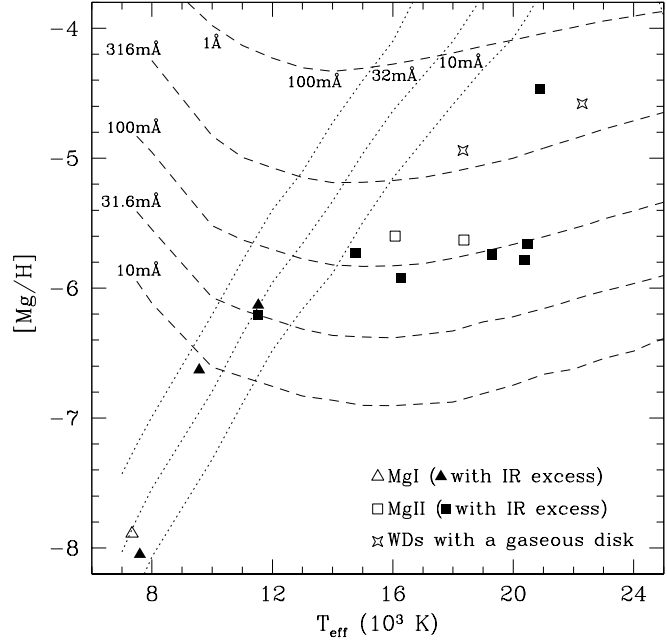}
\caption{Abundance of magnesium versus effective 
temperature. Filled symbols are white dwarfs with confirmed infrared excess, 
and open symbols are white dwarfs either without infrared excess or that have 
not been checked for infrared excess. 
The dashed and dotted lines are curves of constant equivalent 
widths for Mg\,{\sc ii} $\lambda$4481\AA\ and Mg\,{\sc i} $\lambda$5184\AA, respectively.}
\label{fig_abunmg}
\end{figure}

Calcium is the most frequently detected element in DAZ white dwarfs
\citep{zuc2003,koe2005} and is a prevalent metallicity indicator.
Figure~\ref{fig_abunca} shows the calcium abundance versus effective temperature 
measurements in our sample. Stars with confirmed IR excess
indicative of a debris disc are shown with filled squares, and stars that
either do not have IR excess or have not been observed in the mid-IR are
shown with open squares. The calcium abundances in the new cool DAZ 
white dwarfs appear to be above average within their temperature ranges.

The first detection of magnesium in a H-rich white dwarf was in EG102 
\citep{hol1997}. Most white dwarfs with magnesium also show an IR excess.
Magnesium was later detected in several DAZ white dwarfs studied by 
\citet{zuc2003} and in the two white dwarfs (SDSS~J1043+0855 and 
SDSS~J1228+1040) with a gaseous disc \citep{gan2007}. Four of our stars have 
been analyzed by \citet{zuc2003} and our measured magnesium abundances are in 
agreement. In the case of WD2326+049, we measured an
abundance $\log{\rm (Mg/H)} = -6.2$ somewhat lower than their value of -5.8. We 
measured the equivalent widths of Mg\,{\sc i} and Mg\,{\sc ii} lines in UVES and HIRES spectra of WD2326+049 and 
the corresponding abundance varied between -6.0 and -6.25.

Figure~\ref{fig_abunmg} shows the magnesium abundance measurements versus effective temperatures. 
The stars with IR excess are shown with
filled symbols and stars without IR excess with open symbols. The figure also shows
the magnesium abundance determined by \citet{gan2007} for two white dwarfs with a gaseous disc. For WD2346+049, abundance measurements determined from both
Mg\,{\sc i} and Mg\,{\sc ii} lines are shown.

\begin{figure}
\includegraphics[width=0.65\textwidth]{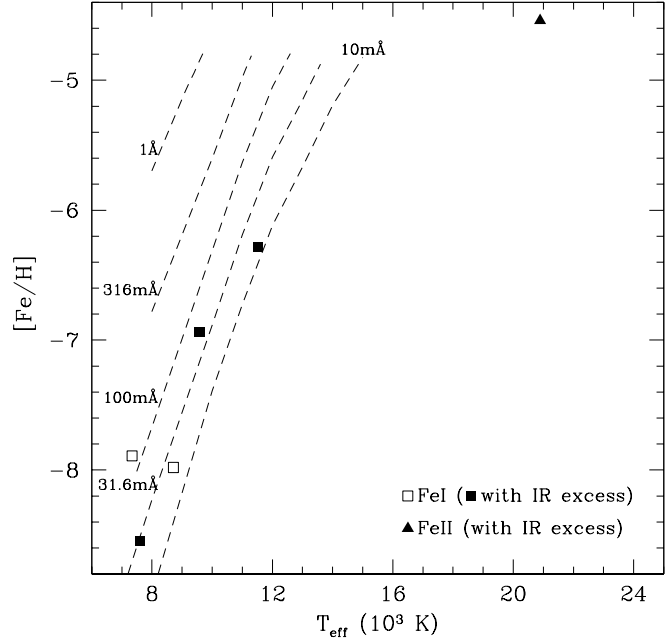}%
\caption{Abundance of iron versus the effective temperature. Filled 
squares are white dwarfs with confirmed infrared excess, and open squares are
white dwarfs either without infrared excess or that have not been checked
for infrared excess. The dashed lines are the equivalent widths for the
Fe\,{\sc i} $\lambda$3581\AA\ line.} 
\label{fig_abunfe}
\end{figure}

We also searched for iron lines and report its detection in five white dwarfs.
Weak Fe\,{\sc i} and Fe\,{\sc ii} lines are difficult to detect and therefore require high
signal-to-noise spectra. Figure~\ref{fig_abunfe} shows the abundance of iron
versus temperature, where all stars except two (WD0208+396, WD1202-232) have confirmed
IR excess (filled symbols). The plot also includes GALEX~J1931+0117 which is
hot enough to display Fe\,{\sc ii} lines (triangle).

Table~\ref{tbl_atm} lists the atmospheric properties and abundances of calcium,
magnesium and iron in the DAZ white dwarfs. The parameters of GALEX~J1931+0117 
are also tabulated.

\begin{figure}
\includegraphics[width=0.65\textwidth]{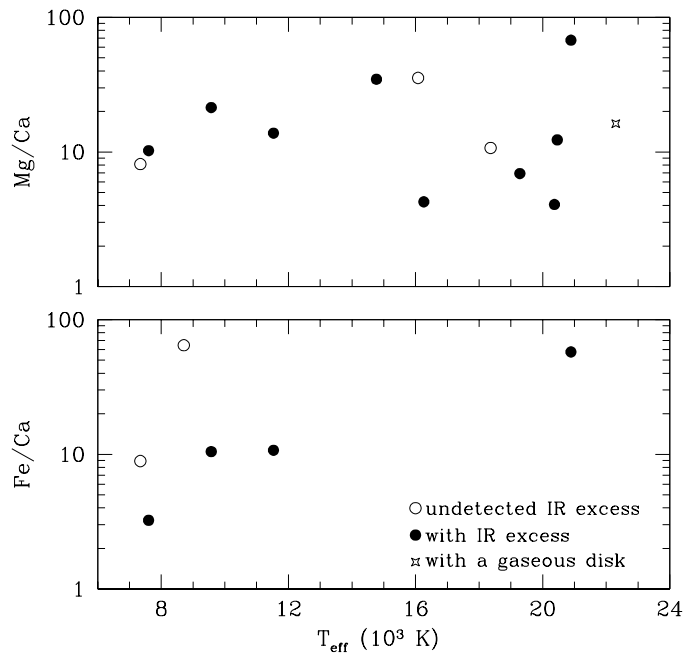}
\caption{
The abundance ratios of magnesium ({\it top}) and iron ({\it bottom}) with 
respect to calcium.}
\label{fig_abun2}
\end{figure}

As first reported by \citet{far2009}, white dwarfs with a higher calcium 
abundance are more likely to harbour a debris disc than those with a lower 
abundance. We also found that the majority of white dwarfs with magnesium ($\approx 75\%$) 
and iron ($\approx 66\%$) in their atmosphere show an IR excess representative of 
a debris disc. Figure~\ref{fig_abun2} shows measured Mg/Ca and Fe/Ca abundance 
ratios. 
The stars without a detectable IR excess were observed with 
Spitzer, but are otherwise similar to the other DAZ white dwarfs. However, their discs may be too narrow to be 
detectable \citep{far2010}.

\begin{table}
\begin{tabular}{lccccc}
\hline
  \tablehead{1}{l}{b}{Name}
  & \tablehead{1}{c}{b}{$T_{\rm eff}$ (K)}
  & \tablehead{1}{c}{b}{$\log{g}$ (c.g.s)}
  & \tablehead{1}{c}{b}{$\log{\rm (Ca/H)}$}  
  & \tablehead{1}{c}{b}{$\log{\rm (Mg/H)}$} 
  & \tablehead{1}{c}{b}{$\log{\rm (Fe/H)}$} \\
\hline
 WD0032-175 &  $9730\pm30$  & $8.11\pm0.03$ & -9.8 & \ldots & \ldots \\
HS0047+1903 & $16080\pm120$ & $7.71\pm0.03$ & -7.2 & -5.6 & \ldots \\
HE0106-3253 & $16260\pm120$ & $7.93\pm0.03$ & -6.6 & -5.9 & \ldots \\
  NLTT06390 &  $6190\pm120$ & $8.39\pm0.25$ & -9.6 & \ldots & \ldots \\
 WD0208+396\tablenote{$T_{\rm eff}$,$\log{g}$ from \citet{gia2005}} &  $7340\pm90$  & $8.10\pm0.04$ & -8.8\tablenote{From \citet{zuc2003}} & -7.9 & -7.9$^{\rm b}$ \\
 WD0243-026 &  $6830\pm70$  & $8.29\pm0.12$ & -9.8 & \ldots & \ldots \\
HS0307+0746 & $10110\pm50$  & $7.97\pm0.04$ & -8.0 & \ldots & \ldots \\
 WD0408-041 & $14770\pm100$ & $7.78\pm0.02$ & -7.3 & -5.7& \ldots  \\
 WD1015+161 & $19280\pm150$ & $7.81\pm0.03$ & -6.6 & -5.7& \ldots  \\
  NLTT23966 &  $7960\pm100$ & $8.02\pm0.14$ & -7.7 & \ldots & \ldots \\
  NLTT25792 &  $8160\pm80$  & $8.12\pm0.11$ & -7.3 & \ldots & \ldots \\
 WD1102-183 &  $8260\pm40$  & $8.17\pm0.06$ & -10.0 &\ldots & \ldots \\
 WD1116+026 & $12450\pm220$ & $7.92\pm0.06$ & -7.7 & \ldots & \ldots \\
 WD1124-293 &  $9470\pm40$  & $8.11\pm0.03$ & -8.8 & \ldots & \ldots \\
 WD1150-153 & $11700\pm110$ & $7.97\pm0.03$ & -7.5 & \ldots & \ldots \\
 WD1202-232 &  $8710\pm40$  & $8.18\pm0.04$ & -9.8 & \ldots & -8.0 \\
 WD1204-136 & $10820\pm60$  & $8.18\pm0.03$ & -8.3 & \ldots & \ldots \\
HE1225+0038 &  $9390\pm40$  & $8.07\pm0.03$ & -9.8 & \ldots & \ldots \\
HE1315-1105 &  $9080\pm40$  & $8.09\pm0.04$ & -9.4 & \ldots & \ldots \\
 WD1337+705 & $19520\pm360$ & $7.87\pm0.06$ & -6.8 & -5.7 & \ldots \\
 WD1455+298\tablenote{$T_{\rm eff}$,$\log{g}$ determined using SDSS spectroscopy.} &  $7600\pm160$ & $8.02\pm0.26$ & -9.1 & -8.0 & -8.6 \\
 WD1457-086 & $20370\pm200$ & $7.80\pm0.03$ & -6.4 & -5.8 & \ldots \\
 WD1614+160 & $16800\pm100$ & $7.80\pm0.02$ & -7.4 &  \ldots& \ldots \\
 WD1826-045 &  $9010\pm30$  & $7.99\pm0.04$ & -9.2 & \ldots & \ldots \\
GALEXJ1931+0117\tablenote{Parameters from \citet{ven2010}} & $20890\pm120$ & $7.90\pm0.06$ & -6.3 & -4.5 & -4.5 \\
 WD2105-820 & $10280\pm50$  & $8.01\pm0.03$ & -8.7 & \ldots & \ldots \\
 WD2115-560 &  $9570\pm40$  & $8.02\pm0.03$ & -8.0 & -6.6 & -6.9 \\
HS2132+0941 & $13060\pm140$ & $7.63\pm0.04$ & -7.9 & \ldots & \ldots \\
 WD2149+021 & $16550\pm70$  & $7.87\pm0.02$ & -8.0 & \ldots & \ldots \\
HE2221-1630 &  $9990\pm40$  & $8.12\pm0.04$ & -8.0 & \ldots & \ldots \\
HS2229+2335 & $18360\pm140$ & $7.84\pm0.03$ & -6.7 & -5.6 & \ldots \\
HE2230-1230 & $19300\pm160$ & $7.73\pm0.03$ & -7.1 & \ldots & \ldots \\
 WD2326+049 & $11530\pm80$  & $8.04\pm0.03$ & -7.3 & -6.2 & -6.3 \\
\hline
\end{tabular}
\caption{Atmospheric properties of selected DAZ}
\label{tbl_atm}
\end{table}

\section{Conclusions}

Spectroscopic studies of accreted material can help reveal the chemical 
composition of the accretion source. In our sample of white dwarfs, magnesium
was detected in 12 white dwarfs (including GALEX~J1931+0117) with temperatures 
ranging from $\approx 7000$ to $20\,000$ K. Iron was detected in five cooler 
white dwarfs ($T_{\rm eff} < 12\,000$ K).

Abundance levels in the atmosphere depend on both the diffusion time-scales of the
element and the accretion rate of individual
elements \citep{koe2006}. 
Although diffusion time-scales are longer in cooler stars, 
the resulting abundances are lower for a given accretion rate because of
the increased mass of the convection zone in such stars.
Accretion rates required to account for the observed abundances of magnesium, calcium,
or iron are $3\times10^8$\,g\,s$^{-1}$ 
for the hottest white dwarf and $3\times10^9$\,g\,s$^{-1}$ for the coolest objects.
Details of these calculations and results will be presented elsewhere.

The signal-to-noise of some of the spectra in this study was too low to detect
any lines of magnesium, iron, or other elements that may possibly be present in the 
white dwarf atmosphere. Higher S/N spectra at high resolution are required 
in order to investigate abundance patterns that may be linked to the
source of the accreted material.

\begin{theacknowledgments}
This research was supported by GA AV grant numbers IAA301630901 and
IAA300030908, and by GA \v{C}R grant number P209/10/0967.
\end{theacknowledgments}

\bibliographystyle{aipproc}   % if natbib is available

\end{document}